\documentclass[reprint,
amsmath,amssymb,aps,prb,floatfix,
]{revtex4-2}
\usepackage[utf8]{inputenc}
\usepackage{amsfonts,amsmath,dsfont}
\usepackage{amssymb, bm, mathrsfs}
\usepackage{graphicx,xcolor,color,soul,natbib,url}
\usepackage{dcolumn}
\usepackage[labelformat=simple]{subcaption}
\usepackage{physics,chemformula}
\usepackage[colorlinks=true,citecolor=blue]{hyperref}
\usepackage[capitalise]{cleveref}
\usepackage{appendix, comment}
\usepackage[mathlines]{lineno}
\usepackage{ragged2e,cancel,relsize}
 
\DeclareCaptionJustification{justified}{\justifying}
\newcommand{\lr}[1]{\left({#1}\right)}
\newcommand{\mbf}[1]{\mathbf{#1}}

\begin{document}
\title{Designer Edge States in Fractional Polarization Insulators}
\author{Wei Jie Chan}
\affiliation{Science, Mathematics and Technology (SMT), Singapore University of Technology and Design (SUTD), 8 Somapah Road, Singapore 487372}
\author{Pei-Hao Fu}
\email{peihao\_fu@sutd.edu.sg}
\affiliation{Science, Mathematics and Technology (SMT), Singapore University of Technology and Design (SUTD), 8 Somapah Road, Singapore 487372}
\author{L. K. Ang}
\affiliation{Science, Mathematics and Technology (SMT), Singapore University of Technology and Design (SUTD), 8 Somapah Road, Singapore 487372}
\author{Yee Sin Ang}
\email{yeesin\_ang@sutd.edu.sg}
\affiliation{Science, Mathematics and Technology (SMT), Singapore University of Technology and Design (SUTD), 8 Somapah Road, Singapore 487372}

\begin{abstract}
We theoretically investigated the topological-protected edge states (TESs) in an anisotropic honeycomb lattice with mirror and chiral symmetries, characterized by an alternative topological invariant - \textit{fractional polarization} (FP), rather than the conventional Chern number.
This system termed an FP insulator is a potential platform for edge-state engineering due to its disconnected TESs.
These disconnected and robust TESs are susceptible to perturbative chiral symmetry-breaking terms which can generate various patterns including the vanishing helical, spin-polarized, and chiral TESs.
Moreover, helical and chiral TES can be achieved by the finite size effect, not possible from the aforementioned terms alone. 
The demonstration of these various TES in an FP-insulator offers an alternative route in designing reconfigurable two-dimensional nanoelectronic devices.
\end{abstract}
\maketitle
\section{Introduction}

Edge states protected by non-trivial bulk topology \cite{Thouless1982} are robust against defects and symmetry-preserved backscattering \cite{Kane2005,Qi2006}.
Usually, the topological-protected edge states (TESs) are guaranteed by the non-vanishing Berry curvature, and their number is related to the Chern number due to the bulk-boundary correspondence \cite{Kane2005,Qi2006}.
Alternatively, the TESs can still be characterized by another bulk topological invariant, i.e. the fractional edge or surface polarization even with a vanishing Chern number \cite{Liu2017,Wu2020}.
This fractional polarization (FP) requires a mirror and chiral symmetry \cite{Benalcazar2017,Ezawa2018} and is related to the quantized Zak phase in time-reversal and inversion symmetric systems \cite{Zak1989,Benalcazar2017}.
Recently, this is commonly realized on a two-dimensional ($2$D) square lattice with anisotropic hopping amplitudes \cite{Benalcazar2017_PRB,Yang2023,Fukui2018,Wienand2022}, $2$D SSH models \cite{Ma2022,Li2022a,Li2022} such as in \ch{A3B} monolayers \cite{Kameda2019} and in anisotropic $2$D honeycomb lattices with a zero Chern number \cite{Ezawa2018,Mondal2022,Mondal2022a}.

\begin{figure}
    \centering
    \includegraphics[width=0.385\textwidth]{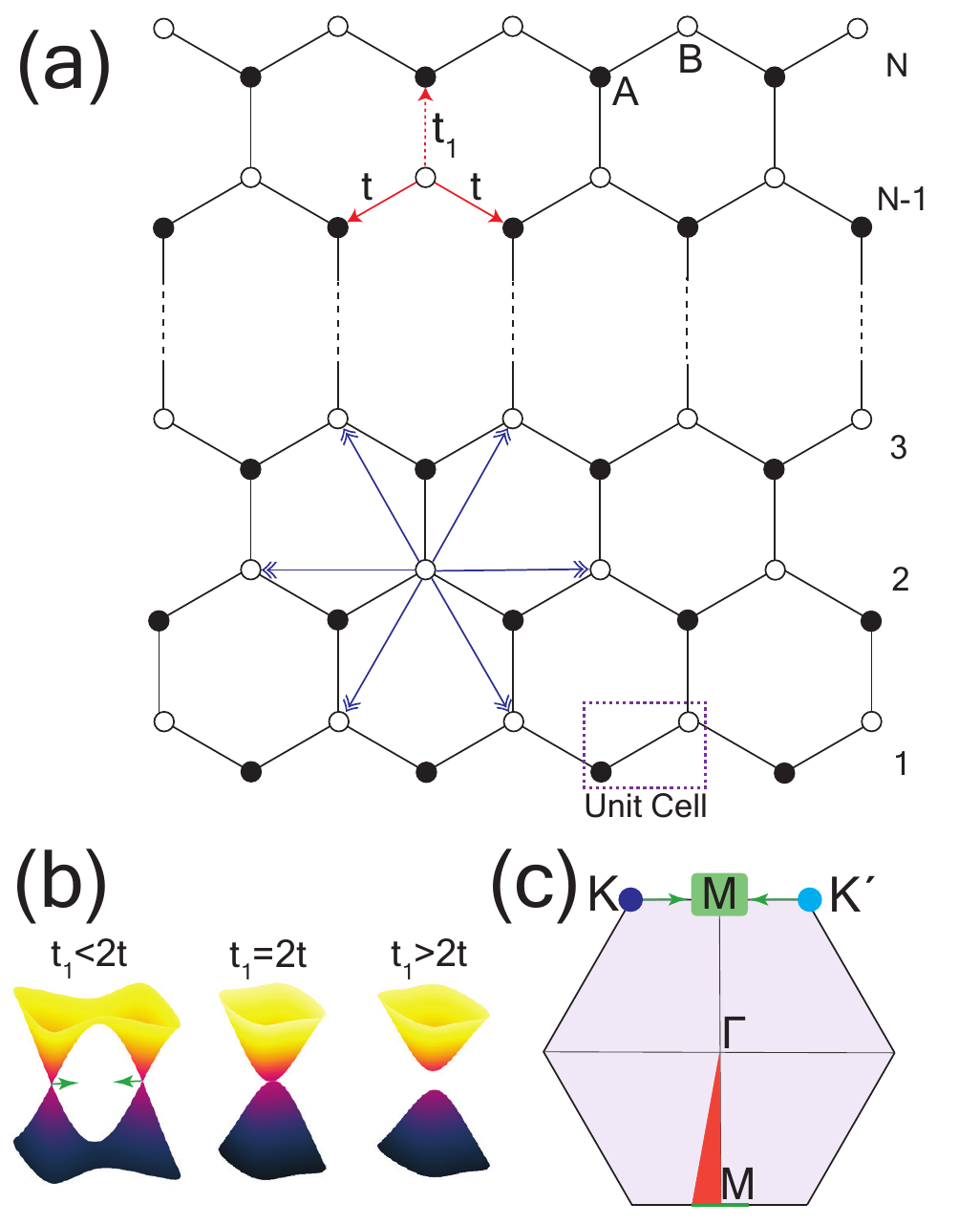}
    \caption{
    A schematic of a semi-Dirac hexagonal lattice model with two sublattices with (next-) nearest-neighbor hoppings denoted by red (blue) arrows in (a). 
    One of the nearest-neighbor hopping terms is anisotropic with $t_1\neq t$.
    The low-energy effective dispersion governed by $t_1/t$ is shown in (b).
    The merging of the $\bm{K}$ and $\bm{K}'$ point into the $\bm{M}$ point is shown within the first Brillouin zone in (c). 
    The fundamental triangle after the merger for $t_1 >2t$ is illustrated as the orange triangle.
    }
    \label{fig:1}
\end{figure}
\begin{figure}
    \centering
    \includegraphics[width=0.43\textwidth]{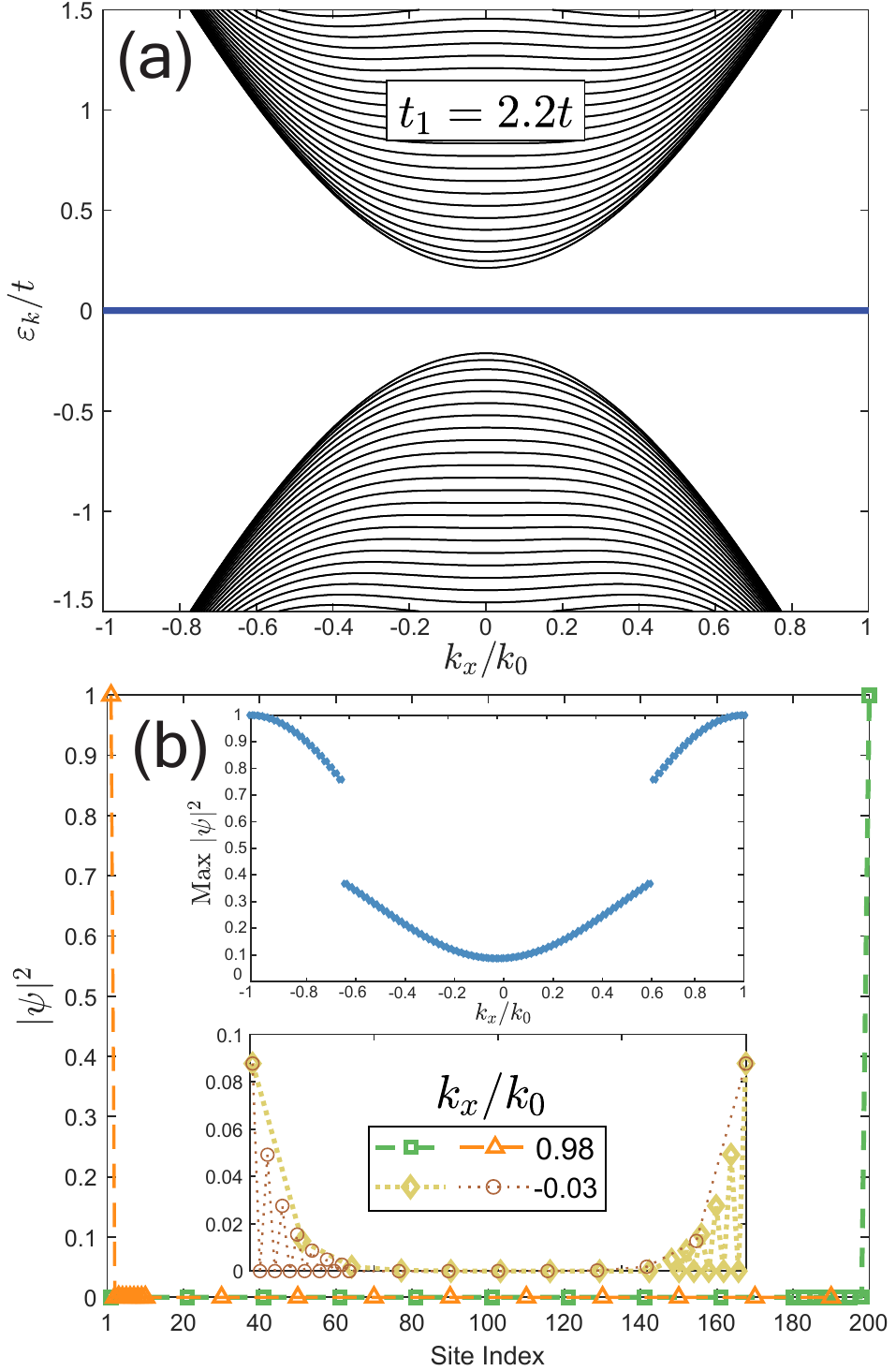}
    \caption{Topological edge states (TESs) isolated from the bulk bands from \cref{eqn:H_edge} with a semi-open boundary condition (sOBC) are plotted with $\varepsilon_{k}/t$ (with subscript $k$ as $k_x$) against $k_x/k_0$ in (a) with $k_0 = \pi/(a\sqrt{3})$ and $t_1=2.2t$ in the presence of NN hoppings only. 
    The total number of sites along the OBC direction ($y$) is $200$ and the value $t_1 = 2.2t$ is fixed for the remainder of the figures.
    The wavefunction probability $\abs{\psi}^2$ against site index at momenta $k_x/k_0$ are shown in figure (b) with its bottom inset.
    The top inset illustrates the regions of localized/delocalized TES across the Brillouin zone (BZ).
    \label{fig:2} 
    }
\end{figure}

Unlike the usual isotropic model in 
the well-celebrated graphene \cite{Mecklenburg2011} and other metamaterials \cite{Qi2021}, the anisotropic model has a modified nearest neighbor (NN) hopping term and is a tunable parameter \cite{Montambaux2009,Montambaux2009_2}.
This tunability merges two Dirac cones in the band structure and results in the topological phase transition [see Fig. \ref{fig:1}(b)] from band-inversion, semi-Dirac, and insulating phase characterized by different transport signatures \cite{Chan2023,Zhai2014} and of potential in designing valleytronics device \cite{Ang2017,Rostamzadeh2022,Choi2021}.
Apart from photonic \cite{Wu2014}, polariton materials \cite{Real2020} and $\alpha$-dice lattice \cite{Carbotte2019,Mandhour2020,Illes2017}, the anisotropic honeycomb lattices are also expected in anisotropic graphene \cite{Delplace2011} and phosphorene \cite{Ezawa2014,Ezawa2018}, monolayer arsenene \cite{Wang2016} and silicene oxide \cite{Zhong2017}.
However, the insulating phase here is not a conventional one, which supports these disconnected TESs \cite{Delplace2011,Ezawa2014,Ezawa2018,Alidoust2020}, and therefore is termed the FP-insulating phase.
Unlike the Zak-insulating anisotropic graphene \cite{Wu2020,Liu2017}, the invariant characterizing the presence of the TES is attributed to the quantized FP in phosphorene due to the mirror symmetries \cite{Ezawa2014}.
Furthermore, the quantization of FP is robust even under perturbative chiral-symmetry breaking terms \cite{Ezawa2018}.
This agrees with \cite{Mondal2022} in the presence of small chiral symmetry-breaking terms in the insulating phase and a vanishing bulk topological invariant.

The aim of this work is to highlight and address the underlying mechanism behind the different possible TES within the FP-insulating phase, which is guaranteed by the mirror symmetries as long as the bulk gap remains open.
This allows us to utilize perturbative chiral (sublattice) symmetry-breaking terms such as spin-orbit coupling (SOC), staggering potential, and exchange proximity effect to realize various TES including the vanishing helical, spin-polarized, and chiral states.
The additional underlying symmetry can also explain the origin of the flat and the two helical TES spectra.
Notably, the finite-size effect can gap out the zero-energy states around the minimal bulk band gap to generate a helical or chiral TES, not possible with the aforementioned chiral symmetry-breaking parameters alone. 

The remainder of this paper is organized as follows.
In \cref{sec:II}, the FP corresponding to the edge state of the Hamiltonian of the anisotropic honeycomb lattice is introduced.
The engineering of the TESs is demonstrated in \cref{sec:III} by considering perturbative chiral symmetry-breaking terms and the finite-size effect.
The conclusion is drawn in \cref{sec:IV}.
\section{Model\label{sec:II}}
\subsection{Tight-Binding Model}
The anisotropic honeycomb lattice in \cref{fig:1}(a) has two nearest neighbors (NN) hopping, $t$ (red arrows), a differing third NN hopping, $t_1$ (red broken arrow) governed by $H_0$, and an inversion symmetry-breaking next-nearest neighbors (NNN) intrinsic SOC contribution term (blue arrows) governed by $H_{\text{SOC}}$ such that
\begin{align}
    H_{0}&= \sum_{\substack{\langle i,j \rangle, s}} t_{ij} c^\dagger_{i,s} c_{j,s}\label{eqn:H0} \\
    H_{\text{SOC}} &=  i\lambda_{\text{SO}}\sum_{\substack{\langle \langle j,j' \rangle\rangle,\\s,s'}} \nu_{jj'}c^\dagger_{j,s} \mbf{s}^{s,s'}_{z} c_{j',s'},\label{eqn:HSOC}
\end{align}
where the hopping term between site $j$ and $j'$ is denoted by $t_{jj'}$.
The SOC coefficient is $\lambda_{\text{SO}}$ with an orientation coefficient $\nu_{jj'}=2(\mbf{d}_{j}\cross\mbf{d}_{j'})/\sqrt{3} =\pm1$, with the unit vector along the bonds between site $j$ and $j'$ as $\mbf{d}_j$ and the spin vector along $z$ between spin $s$ and $s'$ as the spin Pauli matrix $\mbf{s}^{s,s'}_z$.
Additionally, we further include an onsite staggered potential $H_{\text{Stag}}$ and a time-reversal breaking exchange proximity field $H_{\text{Ex}}$ 
\begin{align}
    H_{\text{Stag}} &= \sum_j \lambda_{\Delta,j} c^\dagger_j c_j,\label{eqn:Hstag}\\
    H_{\text{Ex}} &= \sum_{j,s,s'}\lambda_{\text{Ex}} c^\dagger_{j,s} s^{s,s'}_z c_{j,s'},\label{eqn:HEx}
\end{align}
where $\lambda_{\Delta,j} = \pm 1$ denotes A (+) or B (-) sublattice, $\lambda_{\text{Ex},j} = \lambda_{\text{FM}}$ which denotes a ferromagnetic (FM) exchange effect, or antiferromagnetic (AFM), $\lambda_{\text{AFM},j}$ proximity effect.


After a Fourier transformation, the Hamiltonian in momentum space with the basis of ($c_{A,\uparrow},c_{B,\uparrow},c_{A,\downarrow},c_{B,\downarrow}$)$^T$ is (see \cref{appx:bulk} for details) 
\begin{align}
    H_{s_z} &= \begin{bmatrix}
    m_{s_z} + s_z\lambda_{\text{FM}} & f_1-if_2\\
    f_1+if_2& -m_{s_z} + s_z\lambda_{\text{FM}}
    \end{bmatrix},\label{eqn:H_twoband}
\end{align}
where $m_{s_z} = s_z(f_{\text{SO}} + \lambda_{\text{AFM}}) + \lambda_{\Delta}$ and the following terms are 
\begin{align}
    f_1 &= t_1 \cos\lr{k_ya}+2t\cos\lr{\frac{\sqrt{3}}{2}k_xa}\cos\lr{\frac{k_ya}{2}},\label{eqn:Bulkf1}\\
    f_2 &= -t_1\sin\lr{k_ya}+2t\cos\lr{\frac{\sqrt{3}}{2}k_xa}\sin\lr{\frac{k_ya}{2}},\label{eqn:Bulkf2}\\
    f_{\text{SO}} &= -2\lambda_{\text{SO}}\bigg[\sin\lr{\sqrt{3}k_xa}
    \nonumber\\
    &-2\cos\lr{\frac{3k_y a}{2}} \sin\lr{\frac{\sqrt{3}k_x a}{2}}\bigg]. \label{eqn:Bulkfso}
\end{align}
\cref{eqn:H_twoband} can be decomposed into $H_{s_z} = H_0 +H_\mathcal{C}$, where $H_\mathcal{C} = m_{s_z}\sigma_z + s_z \lambda_{\text{FM}} \sigma_0$, consist of all the diagonal terms, with the Pauli matrices denoted as $\sigma_j$, for $j \in \{0,x,y,z\}$, breaks chiral (sublattice) symmetry with $\mathcal{C}=\sigma_z$ and $\mathcal{C}^{-1}H_{s_z}(\bm{k})\mathcal{C} \neq -H_{s_z}(\bm{k})$.
\cref{eqn:H_twoband} obey $\mathcal{M}_{x,y}$ symmetry and it is sufficient to study the TESs from the fundamental triangle, spanning from $k_x/k_0 = 0$ to $-1$ or $1$ within the Brillouin zone (BZ) as highlighted in orange in \cref{fig:1}(c) by considering the underlying mirror symmetry, $\mathcal{M}_{x,y} = i\sigma_{x,y}$ of \cref{eqn:H_twoband},
\begin{align}
    \mathcal{M}^{-1}_{x,y} H_{s_z}(k_x,k_y) \mathcal{M}_{x,y} = H_{s_z}(-k_{x},-k_{y}), \label{eqn:MirrorH}
\end{align}
which plays a pivotal role in the bulk topological invariant of this model.

\begin{figure*}
    \centering
    \includegraphics[width=1\textwidth]{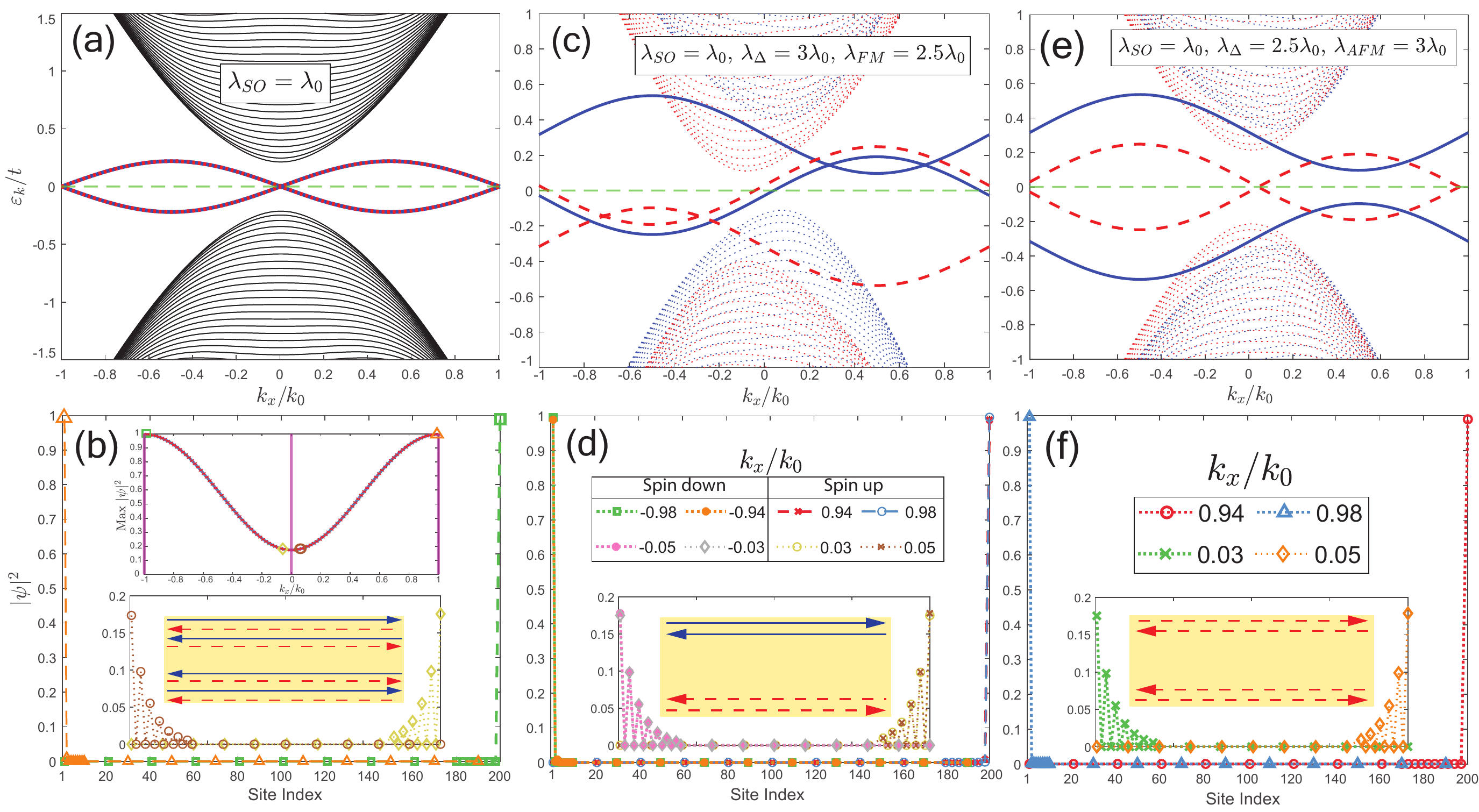}
    \caption{The effects of spin-orbit coupling (SOC) (a), with the addition of an onsite staggering of $\lambda_{\Delta}=3\lambda_0$ with a ferromagnetic (FM) exchange interaction with $\lambda_{\text{FM}} = 2.5\lambda_0$ (c), and the addition of an onsite staggering of $\lambda_{\Delta}=2.5\lambda_0$ with an antiferromagnetic (AFM) exchange interaction with $\lambda_{\text{FM}} = 3\lambda_0$ (e) are shown with $\varepsilon_{k}/t$ against $k_x/k_0$ at $\lambda_{\text{SO}}=\lambda_0$.
    The wavefunction probability is plotted against the site index in the main figure and in the bottom inset with $\abs{\psi}^2$ against the site index for (a), (c), and (e) in (b), (d) and (f) respectively.
    The top inset in (b) illustrates the maximum $\abs{\psi}^2$ against the normalized momenta $k_x/k_0$ with the spin-up(down) indicated as a blue line(red cross), with the purple vertical lines indicating the high symmetric points.
    The yellow schematic in the inset of the bottom figures illustrates the vanishing helical, spin-polarized, and chiral TESs in (b), (d), and (f) respectively.
    }
    \label{fig:3}
\end{figure*}

\subsection{Topological Invariant - Fractional Polarization}
The anisotropic NN hopping, $t_1$ induces a topological phase transition in this model, i.e. from $t_1 < 2t$ through a gapless semi-Dirac phase in $t_1=2t$ to an insulating phase at $t_1>2t$ in \cref{fig:1}(b).
This transition merges the two Dirac cones from the $\bm{K}$ and $\bm{K}'$ valley into a single $\bm{M}$ valley as illustrated in the top half of \cref{fig:1}(c).
This produces an insulating phase with a zero Chern number after the bulk band gaps reopen when $t_1>2t$ \cite{Ezawa2014}.
In the presence of inversion and time-reversal symmetry, this was attributed to the zero Berry curvature $\Omega(k_x,k_y)$, i.e. $\Omega(k_x,k_y) =\Omega(-k_x,-k_y) = 0$ \cite{Liu2017,Wu2020}.
The vanishing of the Chern number with a finite Berry curvature has also been shown in the absence of inversion symmetry \cite{Ezawa2014,Ezawa2018} in phosphorene and anisotropic square lattices \cite{Benalcazar2017}.
However, the existence of isolated topologically protected edge states for $t_1>2t$ indicates that the Chern number is not suitable to describe the bulk topology \cite{Ezawa2014}. 
This was rectified by defining a polarization term \cite{Benalcazar2017} to describe the bulk topology, i.e.
\begin{align}
    P_j = \lr{\int_{\text{BZ}} A_j \text{d}^2\bm{k}} \text{ mod } 1,
    \label{eqn:Polarization}
\end{align}
where the Berry connection is $A_{j,\pm} = -i\mel{\psi_\pm}{\partial_{k_j}}{\psi_\pm}$, with $j$ denotes either the $x$ or $y$ directions for $2$D and the wavefunctions of the two bands are $\psi_\pm$ with $+/-$ denoting the conduction or valence band respectively s.t.
\begin{align}
    \psi_\pm = \begin{bmatrix}
        f_1 - if_2\\ \pm\varepsilon_{\bm{k}}-s_z \lambda_{\text{FM}} - m_{s_z} 
    \end{bmatrix}. \label{eqn:wavefn}
\end{align}

\cref{eqn:Polarization} can still remain quantized to either $0$ or $1/2$,
\begin{align}
    (P_x,P_y) = \lr{0,\frac{1}{2}}, \label{eqn:Py1/2}
\end{align}  
in the absence of inversion symmetry, as long as $\mathcal{M}_{j}$ is present \cite{Benalcazar2017}.

The FP is quantized in the presence of chiral symmetry and was shown to remain well-quantized in the presence of perturbative chiral symmetry-breaking terms \cite{Ezawa2018}.
To ensure the quantization, the terms in $H_\mathcal{C}$ must be small and not be comparable to $t$.
Here, the criteria becomes $\abs{2\lambda_{\text{SO}} + \lambda_{\text{AFM}} + \lambda_{\Delta} + \lambda_{\text{FM}} } \ll \abs{t}$, where the maximum value of $\abs{f_{\text{SO}}}$ is $2\lambda_{\text{SO}}$.
For commonly known $2$D anisotropic honeycomb lattice such as phosphorene \cite{Ezawa2018}, the minimal NN hopping energy is approximately $1.22$eV, which imposes an upper bound of $21\lambda_0$ with $\abs{\lambda_0} = 0.1/\sqrt{3}$.
Hence, the impractically large upper bound demonstrates the perturbative nature of the $H_\mathcal{C}$ and ensures that the bulk band gap remains open and is within the FP-insulating phase.

\begin{figure*}
    \centering
    \includegraphics[width=1\textwidth]{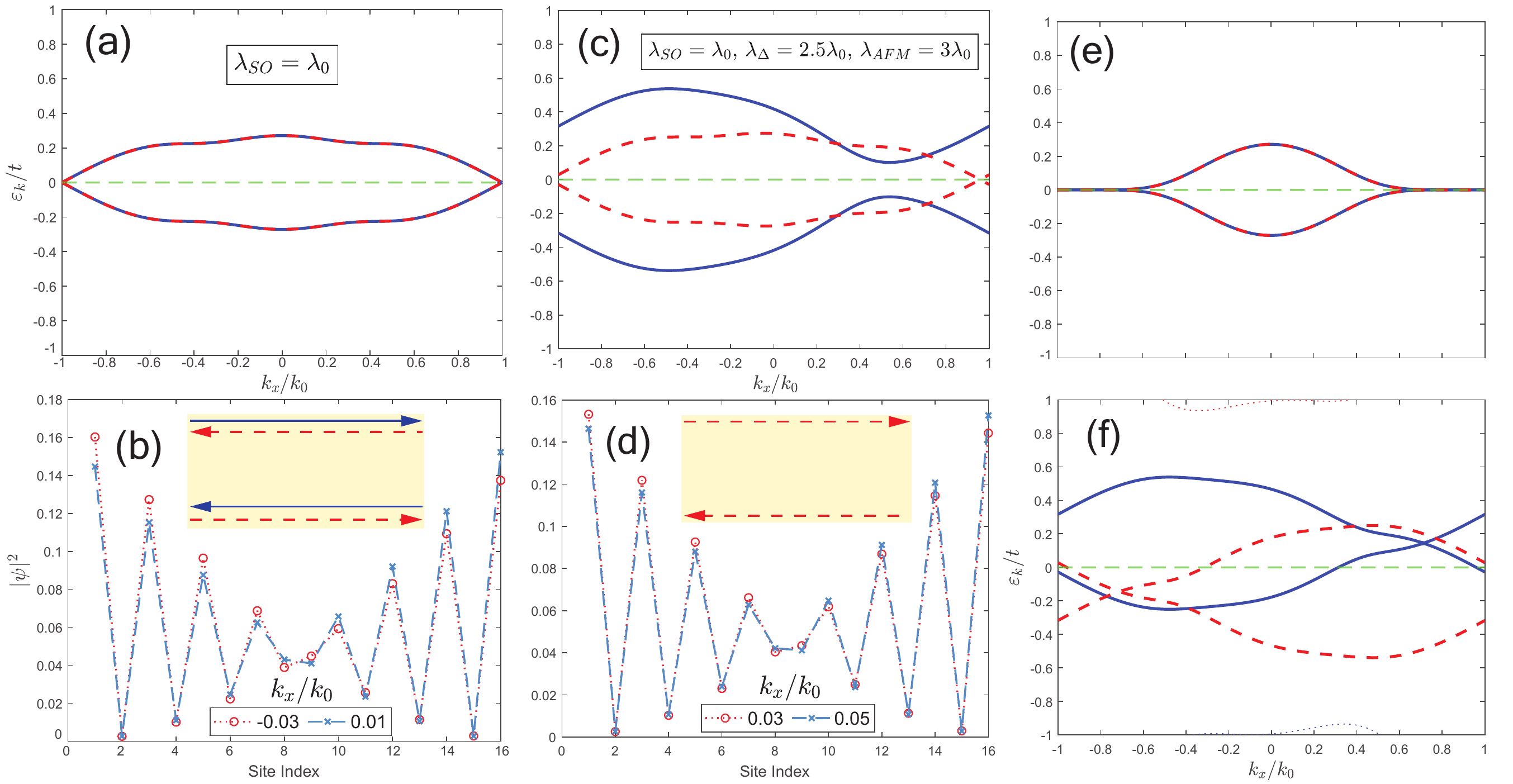}
    \caption{The finite size effect is observed for the TES configurations in \cref{fig:2,fig:3} by gapping the spectrum at $k_x/k_0=0$. 
    This changes the zero-energy modes in (a),(c), and (e) but not in (f) with $\varepsilon_{k}/t$ against $k_x/k_0$.
    The corresponding localization behavior of the gapped is shown in (b) and (d), with the schematic showing the finite size effect induced TES along the boundaries.
    The normalized momenta that are chosen to describe the direction of propagation in the schematic for (b) is $-0.03$ (red circle) and $0.01$ (blue cross), and for (d) is $0.03$ (red circle) and $0.05$ (blue cross).}
    \label{fig:4}
\end{figure*}

\subsection{Edge States}
The TESs in an FP-insulator can be produced by applying a semi-open boundary condition (sOBC) along $y$ s.t. the system becomes a semi-infinite zigzag (ZZ) nanoribbon. 
The corresponding minimal Hamiltonian block derived in \cref{appx:edge} is in the basis of ($c_{1,\uparrow}, c_{2,\uparrow} \dots c_{N,\uparrow}, c_{1,\downarrow} \dots c_{N,\downarrow}$)$^T$ with
\begin{align}
    H^{\text{edge}}_{s_z}(k_x) = &\begin{bmatrix}
        \Tilde{m}(+,+) & f & s_z g_{\text{SO}} &0
        \\ f^* &\Tilde{m}(-,+) & t_1 &-s_z g_{\text{SO}}
        \\s_z g^*_{\text{SO}} & t_1 & \Tilde{m}(+,-) & f
        \\ 0 & -s_z g^*_{\text{SO}} & f^* & \Tilde{m}(-,-)
    \end{bmatrix}
    \label{eqn:H_edge}
\end{align}
where the terms are $\Tilde{m}(\pm,\pm) = \pm m + s_z\lambda_{\text{FM}}$, with $s_z = \pm$ for the FM interaction and
\begin{align}
    m &= s_z (f^{\text{edge}}_{\text{SO}}+\lambda_{AFM})+\lambda_{\Delta}; \label{eqn:effMass}
    \\
    f &= 2t\cos(\sqrt{3} k_x a /2);\label{eqn:fNN}\\ f^{\text{edge}}_{\text{SO}} &= 2\lambda_{\text{SO}} \sin(\sqrt{3}k_xa); \label{eqn:fsoedge}\\
    g_{\text{SO}} &= 2\lambda_{\text{SO}} \sin(\sqrt{3}k_xa/2),\label{eqn:gsoedge} 
\end{align}
with the boundary condition $\psi_{0,B} = \psi_{N+1,A}  = 0$.
Notably, these terms in \cref{eqn:H_edge} are of central importance in manipulating the TES configurations within the FP-insulating phase as discussed below.

\section{Edge States Engineering \label{sec:III}}
The inclusion of the chiral symmetry-breaking terms in $H_\mathcal{C}$ such as SOC, staggering, and exchange proximity effect terms in \cref{eqn:H_edge}, i.e. $f^{\text{edge}}_{\text{SO}}$, $g_{\text{SO}}$, $\lambda_{\Delta}$, $\lambda_{FM}$ and $\lambda_{AFM}$ modifies the behaviors of the TES spectrum while maintaining the quantization of FP as discussed in \cref{sec:II}.
The mechanism behind these modifications can be shown by systematically tracking the zero-energy modes and the underlying symmetries for each newly added term in \cref{eqn:H_edge}.
The order of addition is as follows - NN hoppings, SOC NNN hoppings, staggering with ferromagnetic (FM) proximity effect, and finally swapping the FM with an antiferromagnetic (AFM) interaction. 
Additionally, as the anisotropic NN hopping parameter $t_1$ only alters the bulk band gap, it is fixed at $2.2t$ to remain in the FP-insulating phase.

\subsection{Flat Topological Edge States}
The flat TESs seen from the zero-energy line of \cref{eqn:H_edge} with anisotropic NN hoppings and $m = f^{\text{edge}}_{\text{SO}} = g_{\text{SO}} = 0$, is plotted in \cref{fig:2}(a) with $N = 200$ and $k_x/k_0$ against $\varepsilon_{k}/t$ (with subscript $k$ as $k_x$), where the momenta is normalized by $k_0 = \pi/(a\sqrt{3})$.
The total number of sites is fixed at $N=200$ for both \cref{fig:2,fig:3}.
This originated from the presence of chiral symmetry, $\mathcal{C}=\sigma_z$ which ensures the quantization of the $1$D Zak phase along the ZZ boundaries \cite{Delplace2011}. 
This agrees with \cite{Ezawa2014,Ezawa2018}, in which the flat TES is attributed to the $1$D winding number.
Notably, any $\mathcal{C}$-breaking terms break this quantization, and hence, the spectrum may no longer be flat.

The wavefunction is localized for momenta residing near $k_x/k_0= 1$ and decays into the bulk as momenta decrease towards $0$ as shown in \cref{fig:2}(b) with the wavefunction probability $\abs{\psi}^2$ plotted against the site index. 
Particularly, the orange triangle and green square (yellow diamond and brown circle) lines show the localized (delocalized) TESs with momenta $k_x/k_0 = 0.98$ ($-0.03$.
This is reinforced by seeing that the localized (delocalized) TES resides near the bulk band gap is maximum (minimum) \cite{Zhou2008} in the maximum $\abs{\psi}^2$ against $k_x/k_0$ plot.

Although there is limited tunability in \cref{fig:2} from a fixed \cref{eqn:fNN} and $t_1$, the flat spectrum is a great starting point to observe the effects of SOC, onsite staggering, FM, and AFM exchange interactions.

\subsection{Vanishing Helical Edge States}
The addition of an intrinsic SOC, $\lambda_{\text{SO}}=\lambda_0$ breaks $\mathcal{C}$ and introduced \cref{eqn:fsoedge,eqn:gsoedge} into \cref{eqn:H_edge}.
This modifies the flat spectrum in \cref{fig:2} to produce the vanishing helical TESs at $k_x/k_0 = 0$ and $1$ in \cref{fig:3}(a), as shown by the overlapping spin-up (solid blue lines) and spin-down (dotted red lines) bands.
The zero-energy points are at $k_x/k_0 = 0, 1$ from
\begin{align}
    (\mathcal{M}_x\mathcal{T})^{-1} H^{\text{edge}}_{s_z}(k_x) (\mathcal{M}_x \mathcal{T}) = H^{\text{edge}}_{s_z}(-k_x).
\end{align}
Notably, the time-reversal symmetry that protects the symmetric points with $\mathcal{T} = i\sigma_y \mathcal{K}$ and $\mathcal{K}$ being the complex conjugation operator, is not crucial for the quantization of FP and differs from the usual time-invariant momenta invariant \cite{Kane2005} due to a vanishing $\mathbb{Z}_2$ invariant \cite{Mondal2022}.

These TESs mutually oppose each other and effectively cancel each other along the boundaries as illustrated in the yellow schematic in \cref{fig:3}(b) with blue/red arrows denoted as the spin-up/down TES, in tandem with a vanishing Chern number.
The direction of propagation of TESs is determined by tracking the sign of its velocity, i.e. $v_x\propto \partial \varepsilon_{k} / k_x$ around the zero-energy points with the localized TESs at $k_x/k_0 = -0.98$ (green square) and $0.98$ (orange triangle) in the main figure and delocalized ones at $-0.03$ (yellow diamond) and $0.01$ (brown circle) as shown in the bottom inset.
Also, the top inset shares similar behavior with the top inset in \cref{fig:2}(b) with the spin-up/down indicated by the blue line/red cross with the delocalized/localized states residing near the momenta with the minimal/maximal bulk band gap.
The markers in the top inset correspond to the momenta in \cref{fig:3}(a) with $k_x/k_0 = -0.98$ (green square), $-0.03$ (yellow diamond), $0.01$ (brown circle), and $0.98$ (orange triangle).
Notably, the addition of both onsite staggering potential and the Rashba effect (which breaks in-plane mirror symmetry and hence, the quantization of FP) have been shown to only shift the momenta of the zero-energy modes while retaining the vanishing spin-polarized TESs \cite{Mondal2022}.
Hence, it is natural to add a time-reversal breaking exchange interaction next, to achieve a differing TES configuration while maintaining the quantization of FP.

\subsection{Spin-Polarized Vanishing Edge States}
The inclusion of both the onsite staggering potential, $\lambda_\Delta = 3\lambda_0$ and an FM exchange interaction $\lambda_{\text{FM}} = 2.5\lambda_0$ is shown in \cref{fig:3}(c). 
In comparison to \cref{fig:3}(a), the lifting of $\mathcal{T}$ produces a spectrum that is non-spin-degenerate and has zero-energy modes that slightly deviate from the zero-energy points due to the addition of spin-splitting energy-shifting terms from the combination of both $\lambda_\Delta$ and $\lambda_{\text{FM/AFM}}$.
While spin-splitting is already observed with just $\lambda_\Delta$ in \cref{eqn:effMass} \cite{Mondal2022}, the shifting of the zero-energy modes of each spin band is mainly dependent on $s_z \lambda_{\text{FM}}$, where the energy of the spin bands shift oppositely.

The spin-up/down TESs has positive/negative momenta which produce a vanishing spin-polarized edge state, i.e. opposing spins on opposing boundaries as illustrated in the yellow schematic diagram \cref{fig:3}(d).
This is seen from the localization behaviors of eight momenta values in \cref{fig:3}(d), where the localization behaviors of spin-up bands have momenta at $k_x/k_0 = -0.98$ (green square) and $-0.94$ (orange circle), while delocalized states have momenta of $-0.05$ (pink circle in inset) and $-0.03$ (grey diamond in inset).
Similarly, for the spin-down band, the momenta of the localized and delocalized states are $0.94$ (red cross), $0.98$ (blue circle) and, $0.03$ (yellow circle in inset) and $0.05$ (brown cross in inset) respectively.
Akin to \cref{fig:2}, the delocalized states are located near the minima bulk band gap momenta around $k_x/k_0 = 0$.
Notably, the localization of spin-up(down) states at either $N=200$($N=1$) highlights the presence of spin-polarization, and hence, produces a vanishing spin-polarized edge state.

\subsection{Vanishing Chiral Edge States}
The replacement of the FM interaction with the AFM interaction $\lambda_{\text{AFM}} = 3\lambda_0$ allows the realization of two mutually opposing chiral edge states in \cref{fig:3}(e) at $\lambda_\Delta = 2.5\lambda_0$, and hence, termed a vanishing chiral edge state.
Unlike the FM interaction, the AFM interaction $s_z\lambda_{\text{AFM}}$ acts in tandem with $\lambda_\Delta$, towards the energy of the spin bands.
Hence, this gaps the spin-up (solid blue) bands while keeping the spin-down (dashed red) states intact, realizing a spin-down version of \cref{fig:3}(a).

The vanishing helical edge states in \cref{fig:3}(b) turn into the vanishing chiral edge states as shown in the yellow schematic in \cref{fig:3}(f).
This is seen from the localization behavior of spin-down bands, with the delocalized states at momenta $0.03$ (green cross) and $0.05$ (orange diamond), and localized states at momenta $0.94$ (red circle) and $0.98$ (blue triangle) from \cref{fig:3}(e).

Notably, these three cases have a vanishing TES along the boundaries, in tandem with the vanishing Chern number.
Furthermore, these localization behaviors are consistent regardless of the existence of the chiral symmetry-breaking terms as compared between \cref{fig:2} and \cref{fig:3}, where the delocalized/localized states are near the momenta with the minimum/maximum bulk band gap. 
This points out that the finite size effect can alter the TES, not possible through these chiral symmetry-breaking terms.

\subsection{Finite Size Effect}
The finite size effect is studied by restricting the total site index to $N=16$ in \cref{fig:4} from $N=200$ in \cref{fig:2,fig:3}.
This effect is known to destroy the quantum spin and anomalous hall effect by producing a gap in the TES spectrum \cite{Zhou2008,Fu2014,Liu2016,Ozawa2014} due to the coupling of TES wavefunctions across boundaries as the total $N$ is reduced \cite{Xu2019b}.
However, the size quantization in \cref{fig:4}(a) and (c) produce helical and chiral TESs respectively by gapping out the zero-energy crossing at $k_x/k_0=0$ in \cref{fig:3}(a) and (e), not possible from the addition of the chiral symmetry-breaking terms alone.
The coupling of TESs across boundaries is shown by observing that the wavefunction probability around $k_x/k_0=0$ decays across the site from $N=1$ to $N=16$ in both \cref{fig:4}(b) and (d).
Hence, unlike for the quantum spin and anomalous hall effect \cite{Zhou2008,Fu2014}, the finite size effect creates helical and chiral TESs as shown in the yellow schematic in the inset from the zero-energy modes $k_x/k_0 = \pm 1$.

Similarly, the flat spectrum in \cref{fig:2} produces a gap centered around $k_x/k_0 = 0$ due to the finite size effect as shown in the spectrum in \cref{fig:4}(e).
Finally, the vanishing spin-polarized TES in \cref{fig:3}(c) is shown to be not susceptible to the finite size effect in \cref{fig:4}(f) as the zero-energy crossings are well-localized with its momenta away from the minimum bulk band gap.

\section{Conclusion\label{sec:IV}}
In essence, the mechanism of engineering topological edge states (TESs) is studied within the fractional-polarized (FP)-insulating phase in an anisotropic honeycomb lattice.
This occurs when the anisotropic hopping, $t_1$ exceeds $2t$, producing dispersionless and isolated flat edge bands, with its existence stemming from a zero Chern number, and a quantized fractional polarization term, i.e. $(P_x = 0, P_y = 1/2)$ due to the presence of in-plane mirror and chiral symmetries.
While chiral symmetry-breaking parameters like spin-orbit coupling (SOC), onsite staggering, and the exchange proximity effect, threatens the FP-insulating phase, these terms can be utilized perturbatively to realize different configurations of topological-protected edge states (TESs), namely the vanishing helical, spin-polarized, and chiral TESs.
In particular, this is shown by systematically adding these aforementioned terms while tracking its underlying symmetries and its zero-energy modes.
Due to the existence of delocalized states around the $k_x= 0$ momenta, the finite size effect, which can destroy the quantum spin and anomalous hall effect \cite{Zhou2008,Fu2014}, can be exploited to produce a helical and chiral TES spectrum, not possible through the chiral symmetry-breaking terms.
Finally, the tunability of TESs can provide new avenues in designing electronic devices \cite{Ang2017} and heterojunctions\cite{Betancur-Ocampo2018,Wang2022b}, and also yield unusual behaviors in physical mechanisms such as in electron emission \cite{Ang2018,Ang2021,Chan2022,Luo2023} and transport \cite{Ang2017,Chan2023}.

\section*{Acknowledgment}
Y.S.A. and P.F. are supported by the Ministry of Education, Singapore, under its Academic Research Fund Tier $2$ Award (MOE-T$2$EP$50221$-$0019$). L.K.A. is supported by the ASTAR AME IRG (A$2083$c$0057$).
W.J.C acknowledge the support of MOE PhD RSS. 

\appendix
\onecolumngrid
\section{Full Hamiltonian \label{appx:bulk}}
The bulk energy spectrum of this lattice model is obtained from the addition of \cref{eqn:H0,eqn:HSOC,eqn:Hstag,eqn:HEx} s.t.
\begin{align}
    H_{\bm{k}} =
    \begin{bmatrix}
    f_{\text{SO}} + \lambda_{\text{AFM}} + \lambda_{\Delta} + \lambda_{\text{FM}} & f_1-if_2 & 0 & g_1 - ig_2 \\
    f_1+if_2& -f_{\text{SO}} - \lambda_{\text{AFM}} - \lambda_{\Delta} + \lambda_{\text{FM}} & (g_1+ig_2)^\dagger & 0\\
    0& g_1+ig_2 & -f_{\text{SO}} - \lambda_{\text{AFM}} + \lambda_{\Delta} - \lambda_{\text{FM}} & f_1-if_2\\
    (g_1 - ig_2)^\dagger & 0 & f_1+if_2 & f_{\text{SO}} + \lambda_{\text{AFM}} - \lambda_{\Delta}- \lambda_{\text{FM}}
    \end{bmatrix},\label{eqn:Hfull}
\end{align}
with the following terms in \cref{eqn:Bulkf1,eqn:Bulkf2,eqn:Bulkfso} and
\begin{align}
    g_1 &= i\lambda_\text{R} \lr{e^{ik_ya} - e^{-\frac{ik_ya}{2}}\cos\lr{\frac{\sqrt{3}k_xa}{2}}};\quad 
    g_2= \sqrt{3}\lambda_\text{R}e^{-\frac{ik_ya}{2}}\sin\lr{\frac{\sqrt{3}k_xa}{2}}.
\end{align}
As discussed in \cref{sec:III}, the effect of RSOC is miniature and does not produces a new TESs configuration in agreement with \cite{Mondal2022} for the two helical TESs and hence, can be ignored.
\section{Derivation of Edge States \label{appx:edge}}
The edge states are obtained by applying a semi-open boundary condition (sOBC) along $k_y$ s.t. the system becomes a semi-infinite ZZ nanoribbon.
Next, the NN hoppings from \cref{eqn:Hfull} are decomposed into two hopping directions with respect to the ZZ edge - longitudinal and transverse.
For $l$-th unit cell and width index $m$ as defined in \cref{fig:1}, \cref{eqn:Hfull} can be decomposed into $H_{\text{L}}+H_{\text{T}} + H_{\text{SOC}}$ with
\begin{align*}
    H_{\text{L}}&=t\sum_l \sum_{m=1}^{N} c^\dagger_{l,A}(m) c_{l-1,B}(m) +c^\dagger_{l,B}(m) c_{l,A}(m) + \text{ h.c.}, \quad
    H_{\text{T}}=t_1\sum_l \sum_{m=1}^{N=1} c^\dagger_{l,A}(m+1) c_{l,B}(m) + \text{ h.c.};\\
    H_{\text{SO}}&= i\lambda_{\text{SO}}\sum_{l,s,s'}\nu \Bigg(\sum_{m=1}^{N-1}\bigg[c^\dagger_{l,A}(m+1) \mbf{s}^{s,s'}_z c_{l,A}(m) + c^\dagger_{l+1,A}(m+1) \mbf{s}^{s,s'}_z c_{l,A}(m)\bigg]
    +\sum_{m=1}^{N} c^\dagger_{l+1,A}(m) \mbf{s}^{s,s'}_z c_{l,A}(m)\Bigg)\\ 
    &+ (A\rightarrow B)+ \text{ h.c.} \lr{(m+1\rightarrow m-1) \And (l+1\rightarrow l-1)},
\end{align*}
where $H_{\text{SO}}$ is the intrinsic SOC NNN hopping terms with each line representing different NNN hoppings between sublattice A and B.
This is repeated with its hermitian conjugate terms between width index $m-1$ and $m$, and unit cell $l-1$ and $l$, giving us a total of six NNN terms with its corresponding vector orientation defined as $\nu = \pm 1$.
Additionally, the inclusion of staggering onsite potential, magnetic proximity effect and the RSOC is 
\begin{align*}
    H_{\text{stag}} &= \sum_{l}\sum_{m=1}^N \lambda_{\Delta} c^\dagger_{l,A}(m) c_{l,A}(m) + (A\rightarrow B),\quad
    H_{\text{Ex}} = \sum_{l}\sum_{m=1}^N \lambda_{\text{Ex}} c^\dagger_{l,A}(m) \mbf{s}^{s,s'}_z c_{l,A}(m) + (A\rightarrow B).\\
\end{align*}
where the $\mbf{d}$ vector is similar to $\mbf{d}_{ij}$ in the bulk Hamiltonian.
A block diagonal matrix of size $2N \in 4\mathbb{Z}$, i.e. $H^{\text{edge}}_{2N} = \text{diag}( [H^{\text{edge}}_\uparrow]_{N}, [H^{\text{edge}}_\downarrow]_{N})$ can be obtained by applying a Fourier transform along $k_x$ with periodic boundary conditions (PBC) on the total Hamiltonian define in \cref{sec:II} with $c_{i} = \sum_k 
e^{ik_x\cdot x_{l,(A,B)}} c_{\bm{k}} (m) /\sqrt{L_x}$, $e^{ik_x L_x} = 1$ and $k_x = 2\pi m/L_x $ with $m \in \{0,1, \dots, L_x\}$ and in the basis of ($c_{1,\uparrow}, c_{2,\uparrow} \dots c_{N,\uparrow}, c_{1,\downarrow} \dots c_{N,\downarrow}$)$^T$.
This allows us to get \cref{eqn:H_edge} in \cref{sec:II}.

The set of equations of motion (EOM) can be obtained by inserting a one-particle state \cite{Wakabayashi2010},
\begin{align}
    \ket{\Psi} = \sum_m \lr{\psi_{m,A}c^\dagger_{A,k} (m) + \psi_{m,B}c^\dagger_{A,k} (m)} \ket{0},
\end{align}
where $c^\dagger_{(A,B)} \ket{0} = 0$, with $\ket{0}$ being the vacuum state, into $H\ket{\Psi}=\varepsilon_{k}\ket{\psi}$ s.t.
\begin{align}
    \varepsilon_{k} \psi_{m,A} &= s_0\lr{f_{\text{NN}} \psi_{m,B} + t_1 \psi_{m-1,B} + \lambda_{\Delta} \psi_{m,A}} 
    + s_z\lr{ \lr{f_{\text{SO}} +\lambda_{\text{Ex}}} \psi_{m,A} +g_{\text{SO}}\lr{\psi_{m+1,A}+\psi_{m-1,A}}};\\
    \varepsilon_{k} \psi_{m,B} &= 
    s_0\lr{f_{\text{NN}} \psi_{m,A} + t_1 \psi_{m+1,A} + \lambda_{\Delta} \psi_{m,B}} 
    + s_z\lr{ \lr{f_{\text{SO}} +\lambda_{\text{Ex}}} \psi_{m,B} +g_{\text{SO}}\lr{\psi_{m+1,B}+\psi_{m-1,B}}},
\end{align}
where $\lambda_{\text{Ex}}$ can denote either the FM or AFM exchange.

Hence, the corresponding edge state Hamiltonian in the basis of $\{ \psi_{1,A},\psi_{1,B},\dots,\psi_{N,A},\psi_{N,B}\}$ is shown in \cref{eqn:H_edge}.
The finite termination at the zz edge at width index $0$ and $N+1$ on sublattice B and A respectively impose the boundary condition of $\psi_{0,B} = \psi_{N+1,A} =\psi_{0,A} = \psi_{N+1,B} = 0$ and modifies the EOM s.t.
\begin{align}
    \varepsilon_{k} \psi_{1,A} &= s_0\lr{f_{\text{NN}} \psi_{1,B} + \lambda_{\Delta} \psi_{1,A}} + s_z\lr{ \lr{f_{\text{SO}} +\lambda_{\text{Ex}}} \psi_{1,A} +g_{\text{SO}}\psi_{2,A}};\\
    \varepsilon_{k} \psi_{N,A} &= s_0\lr{f_{\text{NN}} \psi_{N,B} + t_1 \psi_{N-1,B} + \lambda_{\Delta} \psi_{N,A}} 
    + s_z\lr{ \lr{f_{\text{SO}} +\lambda_{\text{Ex}}} \psi_{N,A} +g_{\text{SO}}\psi_{N-1,A}};\\
    \varepsilon_{k} \psi_{1,B} &=s_0\lr{f_{\text{NN}} \psi_{1,A} + t_1 \psi_{2,A} + \lambda_{\Delta} \psi_{1,B}} + s_z\lr{ \lr{f_{\text{SO}} +\lambda_{\text{Ex}}} \psi_{1,B} +g_{\text{SO}}\psi_{2,B}};\\
    \varepsilon_{k} \psi_{N,B} &= s_0\lr{f_{\text{NN}} \psi_{N,A} + \lambda_{\Delta} \psi_{N,A}}+ s_z\lr{ \lr{f_{\text{SO}} +\lambda_{\text{Ex}}} \psi_{N,B} +g_{\text{SO}}\psi_{N-1,B}}.
\end{align}
\twocolumngrid

\end{document}